\documentclass[12pt]{article}
\usepackage{epsf}
                           
\usepackage{amssymb}
\usepackage{amsmath}
\usepackage{color}
\usepackage[utf8]{inputenc}
\usepackage[english]{babel}

\begin{document}
\title*\begin{center}
{\textbf{{\LARGE Asymptotic Regimes of Hadron Scattering in QCD}}}\footnote{The talk at the VIth International conference in memory of A. N. Vasiliev: " Models in Quantum Field Theory". Saint-Petersburg, Peterhof, 27-31 August 2018.} 
\end{center}

\begin{center}
V. A. Petrov\footnote{e-mail: Vladimir.Petrov@ihep.ru} 
\end{center}
\begin{center}
Division of Theoretical Physics, A. A. Logunov Institute for High Energy
Physics,
\begin{center}
NRC "Kurchatov Institute", Protvino, RF
\end{center}
\end{center}

\begin{center}
Abstract
\end{center}
\textit{Possible modes of asymptotic behavior of the amplitudes of hadron scattering in QCD are discussed.
It is shown that the condition of triviality of the scattering matrix when the interaction of the fundamental fields is turned off leads to the impossibility of cross-sections that do not decrease with energy in the case of pure gluodynamics. It is also shown that in the presence of  at least one type of massive quark the ban is lifted. Some consequences of the presence of an infrared fixed point are also discussed.}.

\section*{The Problem} 
This is a commonplace that so far we do not have a full-fledged theory of interaction of hadrons, derived from the first principles of QCD and having a regular way of calculating of hadronic amplitudes, especially at high energies and small momentum transfers. The problem is related to a more general problem that QCD Lagrangian would yield colour confinement with massive colourless physical states (hadrons).

On the other hand, in phenomenological practice we use many general properties of this future theory. For instance, the observed growth of the proton-proton total cross-sections at energies from $ 10 $ to $ 10^{4} $ GeV is generally related with gluon exchanges in the form of colourless Pomeron. There are serious arguments that the nucleon mass is due mainly to its gluon content while quarks play in this respect the minor role.

So, it seems that a pure gluonic colour-confining theory of strong short-range (non zero mass gap) interaction (QGD = "quantum gluodynamics"= quarkless QCD) is quite intelligible and such a theory could, in particular, provide a phenomenon of the ever rising cross-sections. 
 One of the scenarios of the passage from coloured gluon degrees of freedom to the theory of colourless hadrons "glueballs"  could look as follows. Imagine that QGD is resolved, i.e. we have in our disposal the infinite set of gluonic Green functions.
  Confinement means in particular that the usual LSZ procedure, which gives S-matrix elements by a specific amputation of the external legs and taking all external one-particle momenta on-shell,does not work due to uncontrollable IR divergencies.

  So, how should  we proceed? Inevitable action is to consider,
in a generic many-gluon (off-shell) Green function, all possible colourless combinations of external gluon momenta. If theory is non trivial then some combinations should possess pole singularities in corresponding invariant masses which are situated at non zero glueball masses. Taking various manifold residues at all such poles we come to what we can define as hadronic S-matrix. This S-matrix possesses in general both one- and many hadron singularities. The amplitudes are analytic functions of invariant Mandelstam variables as follows from general quantum field principles. No singularities related to gluons appear.

\section*{"Free-field Ansatz"} 

Whatever it is the amplitudes so obtained should contain
in some way the fundamental parameters of the primary QCD Lagrangian. 

We have the following parameters: $ g = \alpha_{s} $, the QCD coupling; $\mu $, the normalization scale.This scale quantifies the ambiguity in multiplication of generalized functions (distributions)which lies in the basis of renormalization. Independence of physical quantities of this mathematical ambiguity results in renormalization invariance, i.e. invariance under the change in 
 $ \mu $ and compensating change in $ g $. Such an invariance for hadronic amplitudes is accomplished by the fact that their dependence on $ g $ and  $ \mu $ is not arbitrary but proceeds via a special combination of these parameters, i.e. via the scale
 \begin{equation}
  \Lambda = \mu \exp \int^{g} dg^{'}/\beta(g^{'})
  \end{equation} 
  where
  \begin{center}
  $ \mu^{2}dg/d\mu^{2} = \beta (g) $.
  \end{center}
Appearance of a massive scale $ \Lambda $ in the primarily massless theory was dubbed in Ref.\cite {col} "dimensional transmutation". This provides an opportunity to get all physical (glueball) states massive. A generic glueball mass 
$ M_{i} $ is proportional to this fundamental scale:
\begin{equation}
M_{i}= c_{i}\Lambda .
\end{equation}
Here $ c_{i} $ are pure dimensionless(generally complex) numbers which do not depend neither on 
$ g $ nor on $ \mu $. They should depend however on the choice of so called  "renormalization scheme". Changing a scheme results in redefining $ g $:
\begin{center}
$\tilde{g} = f(g)$
\end{center}
where $ f $ is an arbitrary regular function constrained by
\begin{equation}
f(0) =0, f^{'}(0)=1  .
\end{equation}
Changing in scheme leads to a multiplicative change of $ \Lambda $ which is to be compensated by the corresponding "opposite" change of $ c_{i} $ because physical masses certainly are independent of any arbitrary scheme change. Possibility of RG invariant scale $ \Lambda $ was discussed in 
\cite{Son}

In these terms  the hadron scattering amplitude $ T(s,t) $ may be cast into the form:
\begin{equation}
T(s,t)= \Phi (\frac{\Lambda^{2}}{s}, \frac{t}{s};{c_{i}})
\end{equation}

 Now, let us take $ g=0 $.
 What can happen in this limit? If we take QGD as a field theory of interacting gluons the limit should bring the Green functions to their free-field values. This seemingly trivial statement leads, nonetheless, to a nontrivial consequences for the \emph{interacting }fields if to take into account that, as follows from Eq.(1),  $ \Lambda^{2}_{\mu fix, g\rightarrow 0}\sim exp (-1/b_{0}g)  $ where $\beta (g)= -b_{0}g^{2}+...,  b_{0} >0   $. In Ref.\cite {west} it was shown that this  "boundary condition" leads in quite a simple way to the explicit form of the asymptotic behaviour of, say, gluon propagator. The reason is that due to RG the dependence on the coupling constant is tightly related with the dependence on momenta.
 Let us try to use the same way for the scattering amplitude of composite particles(hadrons).
 
 Formally we can get four possibilities:
 
 1. The limit $ g \longrightarrow 0 $ just does not exist.
 
 2. The limit is infinite.
  
 3. The limit is some finite function of $s,t$ (including a non zero number 
 
 independent on $ s,t $).
 
 4. The limit is zero.
 
The first item, if accepted, makes any further discussion meaningless. The second item means that hadron interaction gets stronger and stronger as the underlying interaction weakens.
It does not seem very natural. Similarly, the third option looks also physically strange: after switching off the basic, underlying interaction there still remains some non-trivial scattering amplitude of hadrons (which themselves also survive the switch-off).We do not think such a possibility is realistic. In fact, let us suppose that 
\begin{center}
$ \Phi (0, \xi) = 0 , \xi \neq 0$
\end{center}
and let  
\begin{center}
$ \Phi (0, 0) = \infty $,
\end{center}
i.e. we choose the case contradicting to our Ansatz.
This would mean that 
\begin{center}
$\Phi (0, \xi) = \sum_{n=0}^{N} c_{n}\delta^{(n)} (\xi).  $
\end{center}
i.e.
\begin{center}
$ \lim_{s\rightarrow\infty,t/s = \xi } T(s,t) = \sum_{n=0}^{N} c_{n}\delta^{(n)} (\xi).  $
\end{center}
The fourth option is the subject of discussion in the next Section.
\section*{From Zero Coupling to High Energy}
 
Thus, we assume the following "free-field triviality Ansatz":
\begin{equation}
T(s,t)\vert_{g=0} = \Phi (0, \frac{t}{s};{c_{i}}) = 0
\end{equation}
because $\Lambda^{2}_{g\longrightarrow 0} \sim exp (-1/b_{0}g) $.
The limit in (5) is equivalent to the limit (at finite $ g $) $ s\rightarrow \infty , t/s  $ fixed. In other words the hadron scattering amplitude disappears at high energies and fixed angles ($ \cos \theta = 1+2t/s $). This fact is well and for a long time known. For instance, it follows from the "constituent counting rule (CCR)" Ref.\cite{Mat}. We have only to add that our"free-field triviality Ansatz" leads to the \emph{necessary} multiplicative logarithmic correction factors to the CCR power like energy dependence. 
 
Note that our assumption does not exclude the zero angle as well. That is
\begin{equation}
\lim_{g\longrightarrow0} T(s,0)= \lim_{s\longrightarrow \infty}T(s,0)= 0.
\end{equation}
This means, in particular, that
\begin{center}
$  \sigma_{tot}\vert_{s\rightarrow\infty} = 0$

\end{center}
and, what is more, with the rate $ < 1/s $.
Such a behaviour is in blatant contradiction to the observed growth of all measured hadronic cross-sections \cite{pdg} 

But can we consider the  collision energies, achieved by now, as "close to infinity"? So the observed trend formally cannot exclude the disappearing of the total cross section with the energy growth.

However, even if we admit such an evolution we have to admit the existence of some "critical energy" at which the total cross sections will begin to decrease. According to the present data this energy should be knowingly higher than 50 TeV.
With all our knowledge of QCD it seems extremely unlikely that such a scale can arise.

\section*{Forward Scattering and IR Fixed Point} 

There are many evidences that $ \alpha_{s}(0) $ is finite \cite{de}. 
On the other hand, the existence of finite $ \alpha_{s}(0) $
means that beta function posesses an IR fixed point:
\begin{equation}
\beta (\alpha_{s}(0))=\beta (g^{*})= 0
\end{equation}
Often one argues that theory at $g=g^{*}  $ is scale invariant.
To see if and how it works let us consider the gluon propagator in transverse gauge, $ D(q^{2})= \frac{d(q^{2})}{q^{2}}$.
Renormalization group implies the following general form:
\begin{equation}
d(q^{2})= I(q^{2}/\Lambda^{2})\exp (\int^{g} dx \frac{\gamma_{A} (x)}{\beta (x)}) 
\end{equation}
where $ \gamma_{A} (g) $ is the gluon field anomalous dimension while the factor $ I(q^{2}/\Lambda^{2}) $ is invariant under the RG operator  $ \mu \frac{\partial}{\partial \mu} + \beta (g) \frac{\partial}{\partial g} $.
In the vicinity of $ g^{*} $ we have $ \beta (g) \approx \beta^{'} (g^{*})(g-g^{*}),\beta^{'} (g^{*})\equiv \beta^{'}_{*} > 0 $ while
\begin{center}
$ \Lambda^{2}|_{g\rightarrow g^{*}} \sim (g^{*}-g)^{-\frac{1}{\beta^{'}_{*}}}. $
\end{center}

 If we assume that $ d(q^{2})$ is finite at $ g=g^{*} $ then we arrive to the following asymptotic behaviour of the gluon propagator:
\begin{equation}
D(q^{2})\mid _{q^{2}\rightarrow 0} \approx \textit{const}\frac{1}{q^{2}}(\frac{\Lambda^{2}}{q^{2}})^{\gamma_{A}(g^{*})}.
\end{equation}
Note that here $ \Lambda $ is \textit{not} taken at  $g=g^{*}  $. In literature one can meet various options for $ \gamma_{A}(g^{*}) $.  In particular, some authors argue in favour of $ \gamma_{A}(g^{*})=1 $ which would lead to the famous confining potential $ \sim r $ at large distances. Other options extend even to $ \gamma_{A}(g^{*})= -1  $\cite{de} . We are not to continue in this direction more and ask the question of our current interest:
What does it mean for the scattering amplitude?
In asymptotically free theories 
\begin{equation}
\Lambda^{2}_{g=g^{*}} = +\infty.
\end{equation}
As in QGD all hadron masses are proportional to $ \Lambda $ we see that all singularities of the scattering amplitude in this limit are sent to infinity and the very scattering amplitude becomes an entire function of $ s $ and $ t $. Moreover this fact implies that the scattering amplitude 
is a constant number independent of $ s $ and $ t $. Actually this means that\begin{center}
$ T(s,t)\mid_{g=g^{*}} = T(0,0)= const. $
\end{center}

Note that $ T(0,0) $ in the above equation is the value of physical amplitude with no fixing renormalization parameters like $ \alpha_{s} $.
As physical hadrons( glueballs) are assumed to be massive this value is always finite.

So we see that the assumption that the limit $ T(s,t)\mid _{g\rightarrow g^{*}} $ exists is quite consistent with the scale invariance at the IR fixed point though - in distinction with the elementary field propagators -  quite in a trivial way.

\section*{What if to add a single massive quark?} 

We have seen in the previous Sections that our "free-field triviality" Ansatz is hardly compatible with ever growing cross-sections if we deal with quarkless QCD.

Situation drastically changes if we come to quark-gluon QCD.
Here we should take\textit{ massive} quarks because massless quarks lead to hadrons which can be (in spite of confinement) massless, e.g. Goldstone bosons(pions etc).  

Now the QCD Lagrangian contains the term proportional to a new parameter $ m $ of mass dimension 1 (we take for simplicity one single flavour).
This is equivalent to the interaction of this quark with a constant external scalar field $\varphi(x) = m $. Such an interaction induces a new RG invariant , $ M $ . We can choose this second parameter in various ways. Here we take the expression
\begin{equation}
M = m \exp [\int ^{g} dx \gamma_{m}(x)/\beta (x)]
\end{equation}.

This mass scale obeys to the RG equation
\begin{center}
$ (\mu\frac{\partial}{\partial\mu} + \beta (g)\frac{\partial}{\partial g} - \gamma_{m}(g) m \frac{\partial}{\partial m})M =0  $
\end{center}
where $ \gamma_{m}(g) $ is the "mass anomalous dimension".
At small $ g $
\begin{center}
$ M \sim (\frac{1}{g})^{\frac{\gamma^{'}_{m}}{b_{0}}} ,\gamma^{'}_{m}=d\gamma_{m}(g)/dg\mid_{g=0} .$
\end{center}
Now let us ask again what happens with the scattering amplitude $ T(s,0)= F (\frac{\Lambda^{2}}{s}, \frac{\Lambda^{2}}{M^{2}})  $ in the free-field limit?  We get
\begin{center}
$ \lim_{g\rightarrow 0} T(s,0)= F(0,0)  $
\end{center}

Our "free-field triviality" Ansatz gives

\begin{center}
$ F(0,0)=0. $
\end{center}

Is it equivalent to the infinite energy limit as it was in the case of quarkless QCD? 

In no way, as
\begin{center}
$ \lim_{s\rightarrow \infty } T(s,0)= F (0, \frac{\Lambda^{2}}{M^{2}})  $
\end{center}
and we have no reason to ask for this value to be zero.
Thus,the presence of at least one massive flavour opens the way to get ever rising cross-sections.
Just for illustration, we can have the following
expression for high-energy cross-section as 
\begin{center}
$ \sigma_{tot}(s)= \frac{1}{M^{2}} \ln^{2}(s/M^{2})  $
\end{center}
which shows that indefinite rise and free-field triviality are fairly compatible.

\section*{Conclusions} 
 We have shown that the "free-field triviality" Ansatz in quarkless QCD does not admit the ever growing hadronic cross-sections.
 However, the presence of massive quarks changes situation drastically and fairly allows such a growth.
 As the quark masses are stipulated by their coupling to the Higgs field and the value of the Higgs condensate it seems that the rise of the cross-section of \textit{strong} interaction is impossible without mass parameters induced by the \textit{electroweak} sector of the SM.
 It should be noted that the possibility of having non zero "dynamical masses" of quarks in no way devalues our argument, since it relies on Lagrangian mass parameters.
 
\section*{Acknowledgements} 
I am grateful to D. Gross and J. Collins for interesting discussions of the early version of this work 
which is supported by the RFBR Grant 17-02-00120.

\end{document}